\documentclass[12pt]{article}
\usepackage{amsfonts}
\usepackage{amssymb}
\usepackage{amsbsy}
\usepackage{graphics}
\usepackage{epsfig}
\usepackage{color}

\input epsf.sty

\newlength{\TZ}
\newcommand{\BEQ}{\begin{equation}}     
\newcommand{\BEA}{\begin{eqnarray}}
\newcommand{\BD}{\begin{displaymath}}
\newcommand{\EEQ}{\end{equation}}       
\newcommand{\EEA}{\end{eqnarray}}
\newcommand{\ED}{\end{displaymath}}
\newcommand{\bb}{\begin{eqnarray}}
\newcommand{\ee}{\end{eqnarray}}
\newcommand{\nn}{\nonumber}
\newcommand{\e}{{\rm e}}

\newcommand{\D}{{\rm d}}                
\newcommand{\erfc}{{\rm erfc\,}}        
\newcommand{\erf}{{\rm erf\,}}          
\newcommand{\Ai}{{\rm Ai\,}}            
\newcommand{\Bi}{{\rm Bi\,}}
\newcommand{\demi}{\frac{1}{2}}         
\newcommand{\rar}{\rightarrow}          

\newcommand{\appsektion}[1]{\setcounter{equation}{0}\setcounter{subsection}{0}
\section*{Appendix. #1}
\renewcommand{\theequation}{A.\arabic{equation}}
              \renewcommand{\thesection}{A} }

\catcode`\@=11
\def\numberbysection{\@addtoreset{equation}{section}
        \def\theequation{\thesection.\arabic{equation}}}
\numberbysection

\begin{document}

\begin{titlepage}

\vskip 1.5 cm
\begin{center}
{\Large \bf Statistical mechanics of the coagulation-diffusion process with a stochastic reset}
\end{center}

\vskip 2.0 cm
\centerline{{\bf Xavier Durang}$^a$, {\bf Malte Henkel}$^b$ and {\bf Hyunggyu Park}$^a$}
\vskip 0.5 cm
\begin{center}
$^a$School of Physics, Korea Institute for Advanced Study, Seoul 130-722, Korea\\
$^b$Groupe de Physique Statistique,
D\'epartement de Physique de la Mati\`ere et des Mat\'eriaux,
Institut Jean Lamour (CNRS UMR 7198), Universit\'e de Lorraine Nancy,
B.P. 70239, \\ F -- 54506 Vand{\oe}uvre l\`es Nancy Cedex, France\\~\\
\end{center}

\begin{abstract}
The effects of a stochastic reset, to its initial configuration, is studied
in the exactly solvable one-dimensional coagulation-diffusion process.
A finite resetting rate leads to a modified non-equilibrium stationary
state. If in addition the input of particles at a fixed given rate is admitted,
a competition between the resetting and the
input rates leads to a non-trivial behaviour of the particle-density in the stationary state.
{}From the exact inter-particle probability distribution, a simple physical picture emerges: the reset mainly changes
the behaviour at larger distance scales, while at smaller length scales, the non-trivial correlation of the model
without a reset dominates.
\end{abstract}

\vfill

\end{titlepage}

\setcounter{footnote}{0}

\section{Introduction}

Stochastic resets occur quite commonly in very distinct situations.
For example, consider a network of tidal channels on a beach.
{}From time to time, it is washed out by a larger wave.
What would be the average properties of such a network, and how
do they differ close to the water line (when resetting due to waves is frequent), and farther inland ?
Another  often-met instance concerns when searching for some object.
A frequently-used search strategy consists in, after having searched in vain for
some time, to return to the beginning and to start afresh, until the object is found. In remarkable work, Evans and
Majumdar ({\sc em}) \cite{Evans11a} have explored the consequences of stochastic resetting in simple diffusion of
a single random walker. They considered a random walk on the line, with a time-dependent position $x(t)$ and
starting from some initial position $x(0)=x_0\ne 0$, and also with an absorbing trap at the origin $x=0$.
While undergoing the random walk, the particle is reset to its initial position with a rate $r$.
{\sc em} showed that the statistical properties of the random walk are drastically altered by the
resetting. For example, in the long-time limit, the stationary distribution of the particle with reset
is no longer gaussian and the mean time to find a target at the origin becomes finite whenever $r>0$ and actually
has a minimum at some non-trivial value $r^*\ne 0$ \cite{Evans11a}.
Various aspects of searches with reset have been analysed recently
\cite{Abad12,Beni12,Montero12,Mattos12,Sanders12,Arita13} and these considerations have
also been extended to the consideration of teams of
independent researchers \cite{Meija11,Evans11b,Evans12,Franke12}.

Here, we are interested in analysing a simple model of {\em interacting} particles,
subject to a stochastic reset to its
initial configuration. We shall choose here the {\em coagulation-diffusion process},
which in one spatial dimension
is exactly solvable through the method of empty intervals \cite{benA00}
and whose properties have been analysed profoundly in the past, e.g.
\cite{Agha05,benA90,benA98,Burs89,Dahmen95,Doer90,Doer92,Durang10,Durang11,Khorr03,Krebs95,Masser00,Muna06a,Muna06b,Rey97,Spouge88,Racz85}.
The coagulation-diffusion process can be defined in terms of particles which move
diffusively on an infinite chain such that each site
is either empty or occupied by a single particle.
If a particle makes an attempt to jump to a site which is already occupied, it
disappears form the system with probability one, according to $A+A\to A$. As it is well-known,
this system can be exactly solved through the method of empty
intervals, where the central quantity is the probability
$E_n(t)$ that $n$ consecutive sites are empty at time $t$.
The time-dependent
average particle-density is then given by $\rho(t) = \left(1 - E_1(t)\right)/a$,
whereas the $E_n$ satisfy for all $n\geq 1$ the equation \cite{benA90,Burs89,Doer90}
\BEQ \label{1.1}
\partial_t E_n(t) = \frac{2D}{a^2} \left( E_{n-1}(t) + E_{n+1}(t) - 2 E_n(t) \right) \;\; , \;\; E_0(t) =1 \;\; , \;\; E_{\infty}(t)=0
\EEQ
where $a$ is the lattice constant and $D$ the diffusion constant. In the continuum limit, one has
$E_n(t) \to E(t,x)$, the particle-density becomes $\rho(t) = - \left.\partial_x E(t,x)\right|_{x=0}$ and finally
(\ref{1.1}) turns into $\left(\partial_t - 2D \partial_x^2\right)E(t,x)=0$ with the boundary conditions $E(t,0)=1$, $E(t,\infty)=0$.
It has been understood not so long ago how to treat these boundary conditions directly \cite{Durang10}.
The resulting long-time behaviour $\rho(t)\sim t^{-1/2}$
has been confirmed in several experiments involving excitons moving
on polymer chains \cite{Kroon93,Kopelman90} or carbon nanotubes \cite{Russo06,Sriv09}.  The strong
mathematical similarity of these equations with the ones for the probability distribution of a random walker \cite{Evans11a}
initially motivated us to consider a stochastic reset in the coagulation-diffusion process.

We now define the \underline{\em {\bf c}oagulation-{\bf d}iffusion {\bf p}rocess with a stochastic {\bf r}eset} ({\sc cdpr}):
consider a chain with $\cal N$ sites,
each of which is can be occupied by at most one particle. The particles perform random hoppings
to nearest-neighbour sites such that upon encounter of two particles, the arriving
particle disappears. The stochastic reset is described by a given set of probabilities $F_n$
for having $n$ consecutive empty sites.\footnote{For example, for a configuration of uncorrelated particles such
that each single site is occupied with probability $p$, one has $F_n = (1-p)^n$.}
A sweep of the lattice consists of
$\cal N$ steps  of the microscopic dynamics. In each step, one chooses first a particle.
This particle either diffuses with
probability ${\cal P}_{g}=\frac{D}{2D+r/{\cal N}}$ to the left, or else with probability
${\cal P}_{d}={\cal P}_{g}$ to the right
or finally the entire system is reset to the configuration $F_n$ with probability
${\cal P}_r = \frac{r/{\cal N}}{2D+r/{\cal N}}$.
In terms of the empty-interval probabilities $E_n(t)$, the equation of motion (\ref{1.1})
is modified as follows by the stochastic reset
\BEQ \label{1.2}
\hspace{-0.3truecm}\partial_t E_n(t) = \frac{2D}{a^2} \left( E_{n-1}(t) + E_{n+1}(t) - 2 E_n(t) \right) -r E_n(t) + r F_n
\;\; , \;\; E_0(t) =1 \;\; , \;\; E_{\infty}(t)=0
\EEQ
which generalises the problem of the
stochastic resetting of a single random walker as formulated by {\sc em} \cite{Evans11a}.

\begin{figure}[tb]
\centerline{\psfig{figure=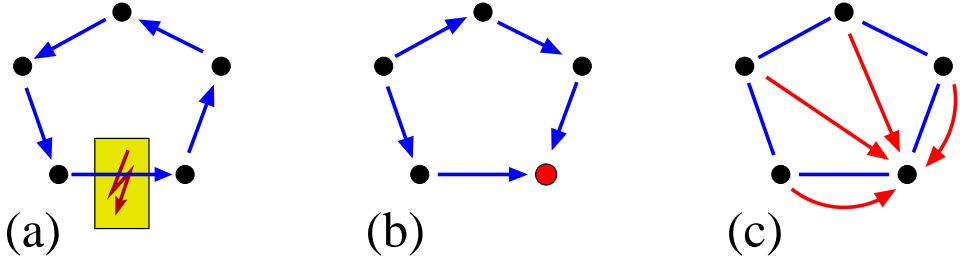,width=5.0in,clip=}}
\caption[fig0]{Schematic illustration of different kinds of non-equilibrium stationary states. The probability currents
are indicated by the arrows. (a) Closed loop
of probability currents, driven through the coupling to external engines. (b) Absorbing stationary state (red dot).
(c) Network of probability currents, modified through a reset to a certain configuration, with the additional probability currents
being indicated by red arrows.
\label{fig0}
}
\end{figure}

Besides being a case study of the influences of a stochastic reset,
the results of this study might be also considered from a different point of view, by studying how a non-equilibrium
system may be set up. A very common way is to allow for non-vanishing {\em probability currents} between the
states of the system (which can be physically realised through coupling to external reservoirs),
leading to closed loops between states, see figure~\ref{fig0}a.
For a detailed review of the properties of the non-equilibrium stationary states, see
\cite{Zia07} and references therein. An alternative possibility occurs for {\em absorbing stationary states},
where the system's evolution goes towards a single absorbing state it cannot leave anymore (figure~\ref{fig0}b),
see e.g. \cite{Henkel09} for
an overview and references therein. Here, and following {\sc em}, we consider what might be happening if the
transition probabilities between states are changed such that the system can return with a certain probability $r$ to its
initial state, see figure~\ref{fig0}c.

This paper is organised as follows. In section~2,
the exact solution of eq.~(\ref{1.2}) is derived. On a discrete chain, a detailed comparison with direct numerical
simulation of the {\sc cdpr} establishes that (\ref{1.2}) is indeed the correct analytical description of the
coagulation-diffusion process with reset. As seen before by {\sc em} for the random walk with reset,
not only the stationary density of single particles and pairs is non-vanishing whenever $r>0$,
but also the entire probability distribution is modified, and this can be illustrated through the
explicit expressions for the $E_n(t)$. As a preparation for later generalisation, the continuum limit of the
same model is derived and leads to the same qualitative conclusions.
In section~3 we extend the model by admitting in addition the possibility
of particle-input on the lattice, at a fixed rate $\lambda$.
It turns out that input and reset interact in a rather non-intuitive way
which leads to a complex and non-monotonous dependence of the stationary particle density on these parameters.
This surprising result will be further illuminated in section~4.
Indeed, the shape of the distribution of the size of the empty intervals between particles
reproduces the non-trivial correlations of coagulation-diffusion
(with or without input) for small intervals
where as the distribution of larger empty intervals is governed by the choice of the reset.
Conclusions are given in section~5. An appendix discusses details of the choice
of the transition rates in Monte Carlo simulations.

\section{Model}

The $1D$ coagulation-diffusion process with a reset ({\sc cdpr}),
as defined in the introduction, can be treated analytically
by introducing the empty-interval probability $E_n(t)$, see \cite{benA00} and refs. therein.
The equation of motion is given by (\ref{1.2}).
In this section, we shall first compute the $E_n(t)$ exactly on an infinite chain and derive from this the
particle- and the pair-densities. A detailed comparison with a direct numerical simulation of the {\sc cdpr}
will illustrate that (\ref{1.2}) gives indeed the correct analytical description. We shall also study the
continuum limit of the model.

An useful alternative route towards the {\em stationary} solution of (\ref{1.2}) starts from the time-dependent
eq. (\ref{1.1}), without a reset. Then the Laplace transform $\overline{E}_n(r):=\int_0^{\infty}\!\D t\, e^{-rt} E_n(t)$,
along with the initial condition $E_n(0)=r F_n$ and with the formal replacement $t\mapsto r$ obeys the stationary
equation (\ref{1.2}) with a vanishing left-hand side, and the boundary condition $\overline{E}_0(r)=r^{-1}$.
Although we shall not follow this route here, this idea might become useful to study the effects of a reset in more general situations.

\subsection{Discrete case}
The solution to (\ref{1.2}) together with the non-trivial boundary condition
$E_0(t)=1$, can be derived by admitting
an analytical continuation to negative indices, via $E_{-n}(t) = 2-E_n(t)$ \cite{Durang10}.
Similarly, we shall define
$F_{-n} := 2 - F_n$ for the resetting distribution.
Then, the generating function
\bb
G(z,t) := \sum_{m=-\infty}^{\infty} z^m \, E_{m}(t)
\ee
obeys, because of (\ref{1.2}), the equation
\bb \label{2.2}
\partial_t G(z,t) = \frac{2D}{a^2}  \left(z+\frac{1}{z} - \left(2+\frac{ra^2}{2D}\right)\right)G(z,t) + rF(z).
\ee
where $F(z)=\sum_{m\in\mathbb{Z}} z^m F_m$ is the generating function of the resetting distribution $F_n$. Eq.~(\ref{2.2}) is
almost automatically solved.
Setting the lattice constant $a=1$ from now on, the full solution can be decomposed as $E_n(t)=E_n^{(1)}(t)+E_n^{(2)}(t)$, with the
`homogeneous' solution ($I_n$ is a modified Bessel function \cite{Abra65})
\bb
E_n^{(1)}(t) = \e^{-(4D+r)t}\sum_{m=-\infty}^{\infty} E_m(0)I_{n-m}(4Dt)
\ee
and the `inhomogeneous' part
\bb
E_n^{(2)}(t) = r\int_0^{t} \D t' \; \e^{-(4D+r)t'}\sum_{m=-\infty}^{\infty} F_m \,I_{n-m}(4Dt').
\ee
Using the analytical continuations $E_{-n}(0)=2-E_n(0)$ and $F_{-n}=2-F_n$, we obtain
\bb \label{En1}
E_n^{(1)}(t) &=& \e^{-(4D+r)t} \left[ \sum_{m=1}^{\infty} E_m(0)\left(I_{n-m}(4Dt)-I_{n+m}(4Dt)\right)
+ 2\sum_{m=1}^{\infty} I_{n+m}(4Dt) + I_{n}(4Dt) \right]
\nn \\
E_n^{(2)}(t) &=& r\int_0^t \D t'\; \e^{-(4D+r)t'}
\left[ \sum_{m=1}^{+\infty} F_m \left(I_{n-m}(4Dt')-I_{n+m}(4Dt')\right) \right. \\ \nn
& &\left. + 2\sum_{m=1}^{\infty} I_{n+m}(4Dt') + I_{n}(4Dt') \right]
\ee
It is straightforward to check that these $E_n(t) = E_n^{(1)}(t) + E_n^{(2)}(t)$
indeed solve the equations of motion and obey the required boundary condition. We also
observe that only the $F_n$ with $n\geq 1$ enter into the final solution.
Since $E_n^{(1)}(t)$ simply reproduces the well-known
solution without a reset \cite{Spouge88,Durang10}, clearly $E_n^{(2)}(t)$ gives
the contributions to the reset. This illustrates how the stochastic reset modifies the entire probability-distribution
of the states in the {\sc cdpr}.

With the knowledge of those empty-interval probabilities,
one can derive the particle-density $\rho(t)=P(\bullet)$
and the pair-density $p(t)=P(\bullet \bullet)$ using the following relations
\bb \nn
\rho(t) &=& 1-E_1(t) \\ \nn
p(t) &=& 1-2E_1(t)+E_2(t).
\ee
The particle-density $\rho(t)$ reads
\bb \label{sol.conc.disc}
\rho(t) &=& \e^{-(4D+r)t}\left(I_0(4Dt)+I_1(4Dt) - \sum_{m=1}^{\infty} \frac{mE_m(0)I_m(4Dt)}{2Dt} \right) \\ \nn
 & & + r \int_0^t \D t' \e^{-(4D+r)t'}\left(I_0(4Dt')+I_1(4Dt')
 - \sum_{m=1}^{\infty} \frac{mF_m I_m(4Dt')}{2Dt'} \right)
\ee
and the pair-density $p(t)$ is given by
\bb
\label{sol.conc.pair}
p(t) &=& \e^{-(4D+r)t} \left[I_0(4Dt)-I_2(4Dt)-2\sum_{m=1}^{\infty} \frac{mE_m(0)I_m(4Dt)}{2Dt} \right. \\ \nn
& &\quad\quad \left. + \sum_{m=1}^{\infty} E_m(0)\left(\frac{(m-1)I_{m-1}(4Dt)}{2Dt}
+ \frac{(m+1)I_{m+1}(4Dt)}{2Dt}\right) \right] \\ \nn
& &+ r \int_0^t \D t' \; \e^{-(4D+r)t'} \left[I_0(4Dt')-I_2(4Dt')-2\sum_{m=1}^{\infty}
\frac{mF_m I_m(4Dt')}{2Dt'} \right. \\ \nn
& & \quad\quad \left. + \sum_{m=1}^{\infty} F_m \left(\frac{(m-1)I_{m-1}(4Dt')}{2Dt'}
+ \frac{(m+1)I_{m+1}(4Dt')}{2Dt'}\right) \right]
\ee

\begin{figure}[tb]
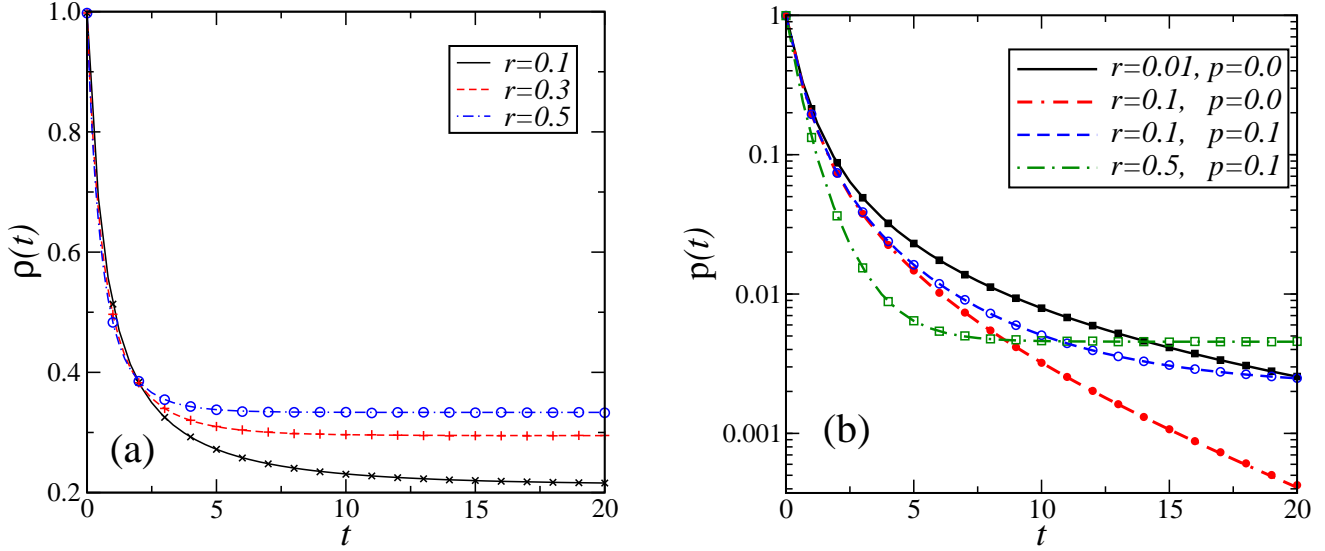

\centerline{\psfig{figure=durang4_reset_conc,width=3.2in,clip=} ~~~~
\psfig{figure=durang4_reset_pair,width=3.3in,clip=}}
\caption[figcp]{Left panel (a): particle-density $\rho(t)$ against time
$t$ for different values of the reset parameter $r$
and particle concentration $p=0.5$.
Right panel (b): pair-density $p(t)$ over against time for various values
of the reset parameter $r$ and the particle concentration $p$.
Here, an uncorrelated reset configuration $F_n =(1-p)^n$ was used.
In both panels, the full line represents the analytic solution
while the symbols show the Monte Carlo simulations.
\label{figcp}
}
\end{figure}

Using several relations from \cite{Abra65}, the stationary
concentration is found from the second term of eq.~(\ref{sol.conc.disc})
\bb \label{2.8}
\rho_{stat,disc}&=& \frac{\sqrt{r(r+8D)} -r}{4D} \\ \nn
& & -\frac{r}{2D}\sum_{m=1}^{\infty}
\left(\frac{1-p}{2}\right)^m \left(\frac{4D}{r+4D}\right)^m
{}_2F_1\left(\frac{m}{2},\frac{m+1}{2};m+1;\left(\frac{4D}{r+4D}\right)^2\right)
\ee

In figure~\ref{figcp}, the analytic results (\ref{sol.conc.disc},\ref{sol.conc.pair})
are compared with simulational results obtained from
the {\sc cdpr} as defined in section~1. Here, a reset to a random uncorrelated configuration of particles of
mean concentration $p$, with an empty-interval distribution $F_n=(1-p)^n$, was used.
We find a clear agreement which
confirms the correctness of the equation of motion (\ref{1.2})
and permits identification with the lattice model.

In the left panel of figure~\ref{figcp}, the relaxation of the particle-density $\rho(t)$ towards its
non-vanishing stationary value (since $p\ne 0$ and $r\ne 0$) is shown. One also sees that the relaxation
towards the stationary value is exponentially fast, instead of the slow algebraic decay $\rho(t)\sim t^{-1/2}$
obtained without reset. In the right panel of figure~\ref{figcp}, an analogous behaviour is found for the
pair-density $p(t)$ when the reset is made to a non-vanishing concentration, $p=0.1$. On the other hand, if
one considers a reset to an empty lattice ($p=0$), one clearly sees that the relaxation has become exponential,
instead of the slow decay $p(t) \sim t^{-1}$ which would hold true in the absence of a reset.

\subsection{Continuum limit}
In the continuum limit, the equation of motion (\ref{1.2}) becomes
\bb \label{eq.comp}
\partial_t E(t,x) = 2D \partial_x^2 E(t,x) - r E(t,x) + r F(x)
\ee
with the boundary conditions $E(t,0)=1$ and $E(t,\infty)=0$. One way to solve this
equation is to separate the empty-interval probability
as $E(t,x) = \frac{1}{2} f(x) + b(t,x)$ such that $f(x)$ will give the stationary solution and
$b(t,x)$ will describe the relaxation towards it. Then, the equation for the stationary probability is
\bb \label{eq.stat}
f''(x) -\alpha^2 f(x) + 2\alpha^2 F(x) = 0 \;\; ; \;\; f(0)=2 \;\; , \;\; f(\infty)=0
\ee
where
\BEQ \label{alpha}
\alpha^2 := \frac{r}{2D}.
\EEQ
The general solution of (\ref{eq.stat}) is readily found by a variation of constants. Taking the two boundary conditions
into account, a straightforward calculation gives
\bb \label{2.12}
f(x) =  2 \e^{-\alpha x}
+ \alpha \int_x^{\infty} \!\!\D x'\: F(x') \e^{\alpha(x-x')} +\alpha \int_0^x \!\!\D x'\: F(x') \e^{\alpha(x'-x)}
-\alpha \int_0^{\infty} \!\!\D x'\: F(x') \e^{-\alpha(x+x')}~~
\ee
To go further, a resetting distribution $F(x)$ must be specified.
In the case of a random distribution where particles has a
probability $p$ to be on a site, the empty-interval probability is given by
$E_n(t)=(1-p)^n$ and, in the continuum limit $x=na$ and
$p\rightarrow0$, the resetting distribution reads $F(x)=\e^{-c x}$, with the reset density
$c=-\ln(1-p) \simeq p +{\rm O}(p^2)$.
Then, the stationary part $f(x)$ is
\bb \label{sol.stat.f}
f(x) = \left(2-\frac{\alpha}{\alpha +c}\right)\e^{-\alpha x} + \frac{\alpha}{\alpha +c}\e^{-cx}
+ \frac{\alpha}{\alpha -c}\left(\e^{-(c-\alpha)x}-1\right)\e^{-\alpha x}
\ee
such that the stationary concentration $\rho_{stat}$ is given by
\bb
\rho_{stat} = \left.-\frac{1}{2}\frac{\partial f(x)}{\partial x}\right|_{x=0} = \frac{\alpha c}{\alpha +c}.
\ee
It can be checked that in the limit $c\ll 1$ of small concentration,
this expression is consistent with the discrete solution (\ref{2.8}).

The dynamical part $b(t,x)$ of (\ref{eq.comp}) satisfies the following equation
\bb
\partial_t b(t,x) = 2D\partial^2_x b(t,x) -r b(t,x)
\ee
with the boundary conditions $b(t,0)=b(t,\infty)=0$. Using a sine Fourier transform,
one can easily find the solution
\bb
b(t,x) = \sqrt{\frac{\pi}{2Dt}\,}\:\e^{-2D\alpha^2t} \int_0^{\infty} \D x' \; b_0(x')
\left[\e^{-\frac{(x-x')^2}{8Dt}}-\e^{-\frac{(x+x')^2}{8Dt}}\right]
\ee
where $b_0(x)$ is the initial condition which can also be decomposed as
$b_0(x) = E_0(x) - \frac{1}{2}f(x)$ where $f(x)$ will give the universal term and
$E_0(x)$ will give the initial-condition-dependent term.
Replacing $f(x)$ by its expression in (\ref{sol.stat.f}), the universal term of the concentration reads
\bb
\rho_{uni}(t) &=&-\left.\frac{\partial b_{uni}(x)}{\partial x}\right|_{x=0} =  \sqrt{\frac{2\pi}{Dt}}
\e^{-2D\alpha^2t} \\ \nn
& & + \frac{2\pi\alpha c}{c^2-\alpha^2}\e^{-2D\alpha^2t}
\left(\alpha\e^{2c^2Dt}\erfc(c\sqrt{2Dt}) - c\e^{2D\alpha^2t}\erfc(\alpha\sqrt{2Dt})\right).
\ee
The initial-condition-dependent term becomes in the special case of initially uncorrelated particles, when $E_0(x)=\e^{-cx}$,
\bb
\rho_{dep}(t)= 2\pi c\e^{-2D(\alpha^2-c^2)t} 
\; \erfc(c\sqrt{2Dt})-\sqrt{\frac{2\pi}{Dt}} \e^{-2D\alpha^2t}
\ee
Hence the full particle-density becomes
\bb
\rho(t) = \rho_{stat} + \rho_{uni}(t) + \rho_{dep}(t) \stackrel{t\to\infty}{\simeq}
\frac{\alpha c}{\alpha + c}
+{\rm O}\left( t^{-1/2} \exp\left(-2D\alpha^2 t\right)\right)
\ee
The introduction of the reset has led to a non-vanishing stationary particle-density
$\rho_{stat}$. For a fixed value of the initial
concentration $c$, $\rho_{stat}$ increases monotonously as a function of the reset rate $\alpha$.
This is qualitatively
analogous to the behaviour of a single random walk with reset \cite{Evans11a}.
Furthermore, the leading approach towards this
new non-equilibrium stationary state is for $\alpha>0$ exponentially fast, but reverts in the
limit $\alpha\to 0$ to the standard slow relaxation
${\rm O}(1/\sqrt{t})$ for the coagulation-diffusion model without reset.

\section{Stochastic reset and input of particles}

We now extend the model and allow the deposition (`input') of particles on the lattice,
with a fixed rate $\lambda$.
Without the reset, this has been treated long ago \cite{Burs89,benA00}.
In view of the technical complexities, we shall
only treat the case of the continuum limit,
where the equation of motion of the empty intervals becomes, for $x\geq 0$
\BEQ \label{3.1}
\partial_t E(t,x) = 2D \partial_x^2 E(t,x) - \lambda x E(t,x) - r E(t,x) + r F(x)
\EEQ
and subject to the boundary and initial conditions
\BEQ \label{3.2}
E(t,0) = 1 \;\; , \;\; E(t,\infty) = 0 \;\; , \;\; E(0,x) = E_0(x)
\EEQ

Again, one may separate this as
$E(t,x) = \frac{1}{2} f(x) + b(t,x)$ such that $f(x)$ will give the stationary solution and
$b(t,x)$ will describe the relaxation towards it. Introducing the abbreviations
\BEQ \label{3.3}
\alpha^2 := \frac{r}{2D} \;\; , \;\; \beta^3 := \frac{\lambda}{2D} \;\; , \;\;
\mu := \frac{\alpha^2}{\beta^3} = \frac{r}{\lambda}
\EEQ
the equation for the stationary empty-interval distribution becomes
\BEQ \label{3.4}
f''(x) - \beta^3 (x+\mu) f(x) + 2\alpha^2 F(x) = 0 \;\; ; \;\; f(0)=2 \;\; , \;\; f(\infty)=0
\EEQ
This may be solved by the standard variation of constants, although the expressions become quite lengthy.
The general solution of the
homogeneous part of (\ref{3.4}) is
\BEQ \label{3.5}
f_{\rm hom}(x) = c_1 \frac{\sqrt{3}}{2} \left(\Bi(\beta(x+\mu)) - \sqrt{3} \Ai(\beta(x+\mu)) \right) +
c_2 \pi \sqrt{3} \Ai(\beta(x+\mu))
\EEQ
where $\Ai$ and $\Bi$ are Airy functions \cite{Abra65}
and $c_{1,2}$ are constants. Then the general solution of (\ref{3.4})
can be written in the form
\BEA
f(x) &=& c_1 \frac{\sqrt{3}}{2} \left(\Bi(\beta(x+\mu))
- \sqrt{3} \Ai(\beta(x+\mu)) \right) + c_2 \pi \sqrt{3} \Ai(\beta(x+\mu))
\nonumber \\
& & + \frac{2\alpha^2}{\beta} \left[\Bi(\beta(x+\mu))
- \sqrt{3} \Ai(\beta(x+\mu)) \right] \int_x^{\infty} \!\D x'\:
F(x') \Ai(\beta(x'+\mu))
\nonumber \\
& & + \frac{2\alpha^2}{\beta}  \Ai(\beta(x+\mu))\int_0^{x} \!\D x'\:
F(x') \left[\Bi(\beta(x'+\mu)) - \sqrt{3} \Ai(\beta(x'+\mu)) \right]
\nonumber
\EEA
Using the asymptotic behaviour of the Airy functions \cite{Abra65},
it is easy to see that $f(\infty)=0$ implies that $c_1=0$ and the second
boundary condition $f(0)=2$ fixes $c_2$. This leads to
\BEA
f(x) &=& \frac{2 \Ai(\beta(x+\mu))}{\Ai(\beta\mu)} - \frac{2\pi\alpha^2}{\beta} \frac{\Bi(\beta\mu)}{\Ai(\beta\mu)}
\Ai(\beta(x+\mu))\int_0^\infty \D x' \; F(x')\Ai(\beta(x'+\mu))
\nonumber
\\
& & +\frac{2\pi\alpha^2}{\beta} \Bi(\beta(x+\mu)) \int_x^\infty \D x' \; F(x')\Ai(\beta(x'+\mu))
\label{eq.f} \\
& & +\frac{2\pi\alpha^2}{\beta} \Ai(\beta(x+\mu)) \int_0^x \D x' \; F(x')\Bi(\beta(x'+\mu))
\nonumber
\EEA

\subsection{Stationary density of particles}
{}From the previous equation (\ref{eq.f}), the stationary density of
particles is obtained, and which can be written down in a scaling form which also
involves the average particle-density $c=-F'(0)$ in the reset configuration $F(x)$, and reads
\BEQ \label{3.7}
\rho_{stat} = - \demi \left.\frac{\partial f(x)}{\partial x}\right|_{x=0}
= c P\left( \frac{c}{\beta}, \beta\mu\right)
\EEQ
with the explicit scaling function
\BEA \label{3.8}
P(u,y) :=  - \frac{1}{u}\frac{\Ai'(y)}{\Ai(y)} - \pi y \left( \Bi'(y)
- \Ai'(y) \frac{\Bi(y)}{\Ai(y)}\right) \int_0^\infty \D Y F(uY/c) \Ai(Y+y)
\EEA
\begin{table}[tb]
\caption[tab1]{Limit behaviour of the scaling function $u P(u,y)$ in eq.~(\ref{3.8}),
for small and large values of $u$ and $y$, respectively. \label{tab1}}
\begin{center}
\begin{tabular}{|l|ll|} \hline
$(2\pi)^{-1} 3^{5/6} \Gamma(2/3)^2 + 3^{2/3} \Gamma(2/3) (4\pi)^{-1} (3\Gamma(2/3)^3 -4\pi^2/3)y$  & $u\to 0$     & $y\to 0$ \\
$y^{-1} -u/c F'(0)$                                                         & $u\to 0$     & $y\to \infty$ \\ \hline
$(2\pi)^{-1} 3^{5/6} \Gamma(2/3)^2 + (2\pi)^{-2}3^{5/3}\Gamma(2/3)^3 y$     & $u\to\infty$ & $y\to 0$ \\
$y^{1/2}$                                                                   & $u\to\infty$ & $y\to \infty$ \\ \hline
\end{tabular} \end{center}
\end{table}
%
\noindent
Herein, the first scaling variable $u=c/\beta$
measures the ratio of the particle-density of
the reset configuration with respect to the stationary density without reset and  the second scaling variable
$y=\beta\mu=(\alpha/\beta)^2$ is a function of the ratio of the reset rate with the input rate. The scaling
function $P=\rho_{stat}/c$ itself measures directly
the stationary density in units of the reset density $c$.
In table~\ref{tab1} the
asymptotic behaviour of the scaling functions for $u$ and $y$
small or large are listed (remarkably, the limits are independent
of the choice for the reset distribution $F(x)$). For $y\to 0$ one always recovers the
known stationary particle density of the case without reset \cite{Burs89,benA00}, as expected.
However, the qualitative behaviour of $P(u,y)$ as a function of
$y$ changes according to the fixed value of $u$. When
$u\ll 1$, $P(u,y)$ will monotonously decrease as a function of $y$,
whereas for $u\gg 1$, one observes a monotonous
increase. From table~\ref{tab1}, this can be read off analytically from the small-$y$ behaviour
\BEA
u P(u,y) \simeq \left\{ \begin{array}{ll}
0.729 - 0.531\, y & \mbox{\rm ~~;~ if $u\to 0$} \\
0.729 + 0.392\, y & \mbox{\rm ~~;~ if $u\to \infty$}
\end{array} \right.
\EEA
\begin{figure}[tb]
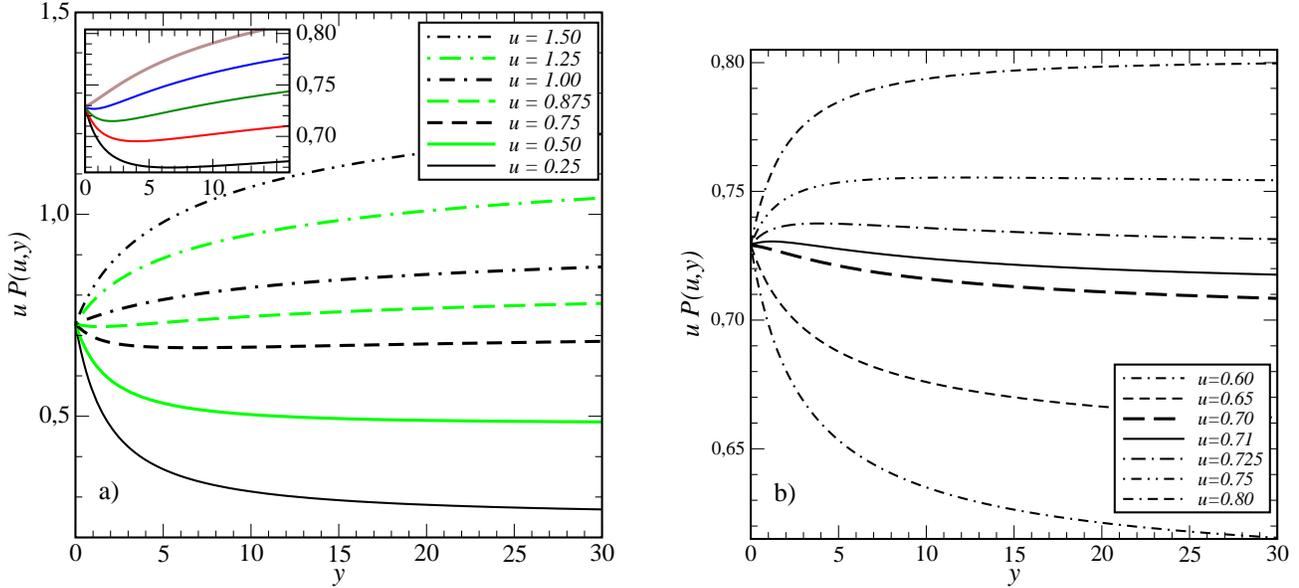

\centerline{\psfig{figure=durang4_Reset_fig1a,width=3.2in,clip=} ~~~~
\psfig{figure=durang4_Reset_fig1b,width=3.2in,clip=}}
\caption[fig1]{Plot of the scaling function $u P(u,y)$ as a function of $y$ and for different values of
$u$. The left panel (a) uses the reset function
$F(x)=\exp(-cx)$, appropriate for uncorrelated particles; the inset shows the same plot for
the values $u=[0.75,0.80,0.85,0.90,0.95]$ from bottom to top. In the right panel (b), the scaling function
$u P(u,y)$ for the choice
$F(x)=\erfc( \frac{1}{2} \sqrt{\pi\,}\, c x)$ is shown for comparison. \label{fig1}
}
\end{figure}

The surprisingly complex behaviour of $P(u,y)$ is
further illustrated in figure~\ref{fig1} and also depends in a subtle way
on the choice of the resetting configuration $F(x)$. We begin with the case
$F(x)=e^{-cx}$ of uncorrelated particles with
concentration $c$. From figure~\ref{fig1}a, it
can be seen that there is also an intermediate range $u\approx 0.8 - 0.9$,
when $P(u,y)$ is a non-monotonous function of $y$.
{}From the inset in figure~\ref{fig1}a, it can be seen that $P(u,y)$
goes through a minimum before the final growing regime for
$y$ sufficiently large is reached.
A local analysis shows that $\partial P(u,y)/\partial y<0$ for $u=u_c:=\leq 0.9295765\ldots$, which means
that this minimum exists for all $u < u_c$.

For different choices of $F(x)$, one may encounter different scenarios.
In figure~\ref{fig1}b, we show results for the choice
$F(x) = \erfc (\demi \sqrt{\pi\,}\, cx)$.\footnote{In the simple coagulation-diffusion process starting from an initially  fully
occupied lattice, this is the exact shape of
the empty-interval probability $E_n(t)$ with a known time-dependent density $c=c(t)$ \cite{Durang10}.}
Without neither a reset nor an input,
the system will converge towards this distribution,
with a certain (time-dependent) particle-density $c$. While for
$u\ll 1$ and for $u\gg 1$, the qualitative behaviour is analogous
to the one seen before, a non-monotonous behaviour now occurs in the middle region $u\approx 0.7 - 0.75$,
with $u_c=0.7010036\ldots$. However, for $u$ not too far above $u_c$,
the scaling function now rather shows a maximum.

\begin{figure}[tb]
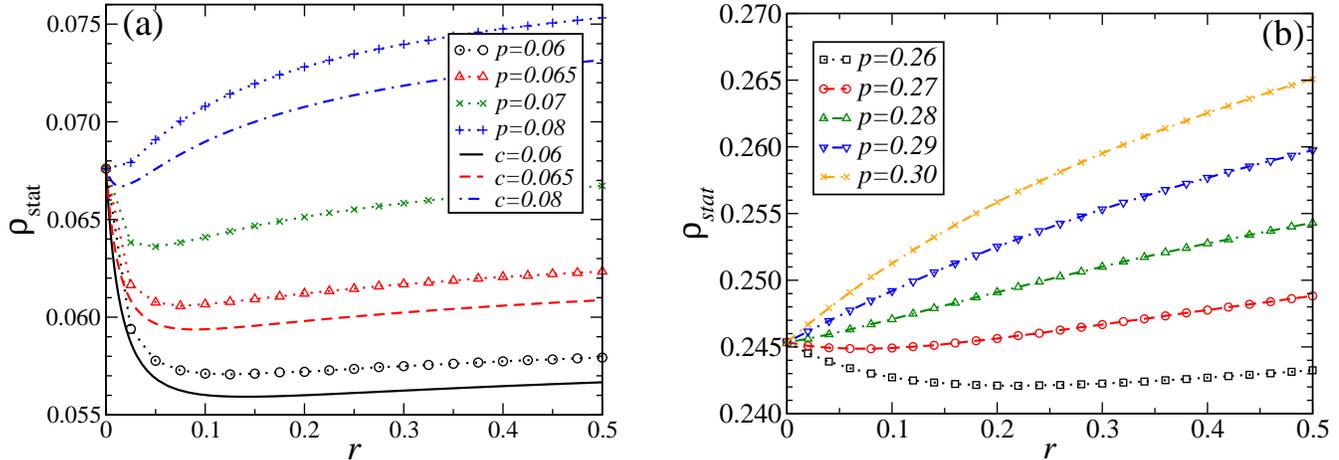

\centerline{\psfig{figure=durang4_reset_Cstat_L0008,width=3.2in,clip=} ~~~~
\psfig{figure=durang4_reset_Cstat_L004,width=3.4in,clip=}}
\caption[fig1]{Plot of the stationary probability $\rho_{stat}$ as a
function of the reset parameter $r$ towards uncorrelated particles for two values of the input parameter:
(a) left panel $\lambda =0.0008$, (b) right panel $\lambda=0.04$.
In the left panel, the full lines without symbols give the analytical
solution (\ref{3.7}) in the continuum limit.
\label{fig2}
}
\end{figure}

The analytical results of this section were obtained in the continuum limit.
In figure~\ref{fig2}, results from a direct numerical simulation of the {\sc cdpr} with input are shown, for
two values of the input rate $\lambda$, and in the case of a reset towards uncorrelated particles.
Qualitatively, the behaviour of the stationary density is analogous to the one seen in figure~\ref{fig1}a and hence is
in qualitative agreement with the analytic solution (\ref{3.7}), obtained in the
continuum limit. While both the discrete and the continuum versions of the {\sc cdpr} lead to the same qualitative conclusions,
the precise form of the stationary density is influenced by the fact that the simulations were carried out on
a discrete chain.

\subsection{Dynamics}

Now, we complete this study by deriving the analytical
solution for dynamical part $b(t,x)$ which satisfies
\BEQ
\partial_{\tau} b(\tau,x) = \partial_x^2 b(\tau,x) - \beta^3 x b(\tau,x) - \alpha^2 b(\tau,x)
\EEQ
where time was rescaled according to $\tau := 2D t$ and one also has the boundary and initial conditions
\BEQ
b(\tau,0) = b(\tau,\infty) = 0 \;\; , \;\; b(0,x) = b_0(x) = 2 E_0(x) - f(x)
\EEQ
In principle, one may solve this by using a Laplace transform
$\bar{b}(s,x) = \int_0^{\infty} e^{-s\tau} b(\tau,x)$. This gives
the equation
\BEQ
\partial_x^2 \bar{b}(s,x) - \beta^3 x \bar{b}(s,x) - (s+\alpha^2) \bar{b}(s,x) + b_0(x) = 0
\EEQ
along with the boundary conditions $\bar{b}(s,0)=\bar{b}(s,\infty)=0$.
This is the same type as eq.~(\ref{3.4}). An analogous
straightforward, but just a little tedious, calculation leads to
\typeout{*** saut de page ***}
\newpage
{\small \BEA
\lefteqn{ \bar{b}(s,x) = \frac{\pi}{\beta}
\left\{ -\int_0^{\infty} \!\D Y\: b_0(Y) \Ai\left(\beta(Y+\mu)+s/\beta^2\right)\Ai\left(\beta(x+\mu)+s/\beta^2\right)
\frac{\Bi(\beta\mu+s/\beta^2)}{\Ai(\beta\mu+s/\beta^2)} \right.}
 \\
&\!\!\!\!\!\!+& \!\!\!\!\!\!\left.
\int_x^{\infty} \!\D Y\: \Ai\left(\beta(Y+\mu)+s/\beta^2\right)\Bi\left(\beta(x+\mu)+s/\beta^2\right)
+\int_0^{x} \!\D Y\:\Bi\left(\beta(Y+\mu)+s/\beta^2\right)\Ai\left(\beta(x+\mu)+s/\beta^2\right) \right\}
\nonumber
\EEA
}
Generalising \cite{Rey97}, the inverse Laplace transform is now formally found from the poles of $\bar{b}(s,x)$,
which arise via the zeroes of the Airy function, in the first term. The result is
\bb
b(t,x) = -\pi\beta \sum_{n=1}^{\infty} \int_0^{\infty} \D x'\: b(0,x') \Ai(\beta x' +a_n)\Ai(\beta x +a_n)
\frac{\Bi(a_n)}{\Ai'(a_n)}\exp(-t(r+|a_n|\beta^2)
\ee
where $a_n$ is the $n^{\rm th}$ zero of the Airy function \cite{Abra65}.
{}From this, one can read off the leading relaxation time  $\tau_{\rm rel} = 1/(|a_1|\beta^2 + r)$, which is finite.

\section{Inter-particle distribution function}

In order to better appreciate the physical nature of the stationary state, we now study the properties of the
{\it {\bf i}nter-{\bf p}article {\bf d}istribution {\bf f}unction} ({\sc ipdf}), denoted here as ${\cal D}(x)$. On
a discrete lattice ${\cal D}_n$ would be the probability that the nearest neighbour of a particle would be at a distance of
$n$ sites. In the continuum limit, this becomes ${\cal D}(x)$.
The relation to the stationary empty-interval probability $E_{stat}(x)=\demi f(x)$ is well-known \cite{benA00}
\BEQ \label{4.1}
{\cal D}(x) = \frac{1}{2\rho_{stat}}\:\frac{\partial^2 f(x)}{\partial x^2}
\EEQ
with the stationary density $\rho_{stat}$ found above.
For the following discussion, we shall require the well-known expressions for ${\cal D}(x)$, as listed in table~\ref{tab2},
for three paradigmatic systems; see e.g. \cite{benA00} and references therein for the computational details.

\begin{table}[tb]
\caption[tab2]{The stationary empty-interval probability $E_{stat}(x)$
and the corresponding {\sc ipdf} ${\cal D}(x)$ for three types
of systems: (a) uncorrelated particles, (b) coagulation-diffusion and (c) with additional particle input.
The distributions are characterised by the model parameters $c$ and $\beta$. \label{tab2}}
\begin{center}
\begin{tabular}{|cl|ll|} \hline
    & type                  & $E_{stat}(x)$               & ${\cal D}(x)$ \\ \hline
(a) & uncorrelated          & $\exp(-c x)$                & $c\exp(-cx)$ \\
(b) & coagulation-diffusion & $\erfc(\demi\sqrt{\pi\,} c x)$ & $\demi\pi c^2\, x \exp\left(-\frac{\pi}{4} c^2 x^2\right)$ \\
(c) & with particle-input  & $\Ai(\beta x)/\Ai(0)$       & $\beta^2 x \Ai(\beta x)/|\Ai'(0)|$ \\ \hline
\end{tabular} \end{center}
\end{table}

Clearly, the case (a) of uncorrelated particles will be an important test case for the study of the effects of a reset.
Recall that this distribution is also obtained for a {\em reversible} coagulation-diffusion process with the
extra reaction $A\to A+A$ \cite{benA00}, such that the stationary state is an equilibrium state. The second
case (b) describes the correlations spontaneously generated during a coagulation-diffusion process $A+A\to A$.
In these two cases, the parameter $c$ denotes the average particle-density. It is known
that for an arbitrary initial condition in pure coagulation-diffusion,
the system converges towards this distribution, with an explicitly known time-dependent
concentration $c=c(t)$ \cite{Spouge88,benA00,Durang10}.
Finally, case (c) gives the stationary distribution with an additional input of particles.
For later comparisons, recall the asymptotic form
${\cal D}(x) \stackrel{x\to\infty}{\sim} x^{3/4} \exp\left(-\frac{2}{3}(\beta x)^{3/2}\right)$.

It is clear from table~\ref{tab2} that the three systems are clearly
distinguished via their {\sc ipdf}s for large interval
sizes $x\to\infty$. This observation will become the central tool in our analysis of the {\sc ipdf} with a reset.

\subsection{{\sc ipdf} without input}

Using the previous expression (\ref{2.12}) of the function $f(x)$, and the definition (\ref{4.1}), the
{\sc ipdf} can be cast into a scaling form
\BEQ
{\cal D}(x) = \alpha D(\xi,v) \;\; , \;\; \xi := c x \;\; , \;\; v := \frac{\alpha}{c}
\EEQ
with the explicit scaling function
\BEA
\lefteqn{D(\xi,v) = \frac{\alpha}{\rho_{stat}} \left\{ e^{-v\xi} - F\left(\frac{\xi}{c}\right) \right.} \\
&+&\left. \frac{v}{2} \left[ \int_0^{\xi}\!\D Y\: F\left(\frac{Y}{c}\right) \, e^{v(Y-\xi)}
+ \int_{\xi}^{\infty} \!\D Y\: F\left(\frac{Y}{c}\right) \, e^{v(\xi-Y)}
- \int_0^{\infty}\!\D Y\: F\left(\frac{Y}{c}\right) \, e^{-v(Y+\xi)}
\right] \right\}
\nonumber
\EEA
In what follows, we shall discuss two specific examples:
first, for a reset to uncorrelated particles with mean density $c$,
one has $F(x)=e^{-cx}=e^{-\xi}$ and
\bb \label{4.4}
D_{(a)}(\xi,v) = \, \frac{\exp(-v\xi)-\exp(-\xi)}{1-v}
\ee
Second, for a reset to a coagulation-diffusion configuration with density $c$, one has
$F(x) = \erfc(\demi\sqrt{\pi}\,\xi)$, hence
{\small\bb \nn
D_{(b)}(\xi,v) &=& \frac{1}{2 \erfc\left(v/\sqrt{\pi}\right)} \left[ e^{-\xi v}\left(1+\erf\left(\xi\frac{\sqrt{\pi}}{2}
-\frac{v}{\sqrt{\pi}}\right)\right)
-e^{ \xi v}\erfc\left(\xi\frac{\sqrt{\pi}}{2}+\frac{v}{\sqrt{\pi}}\right)\right]
\label{4.5}
\ee}
\begin{figure}[tb]
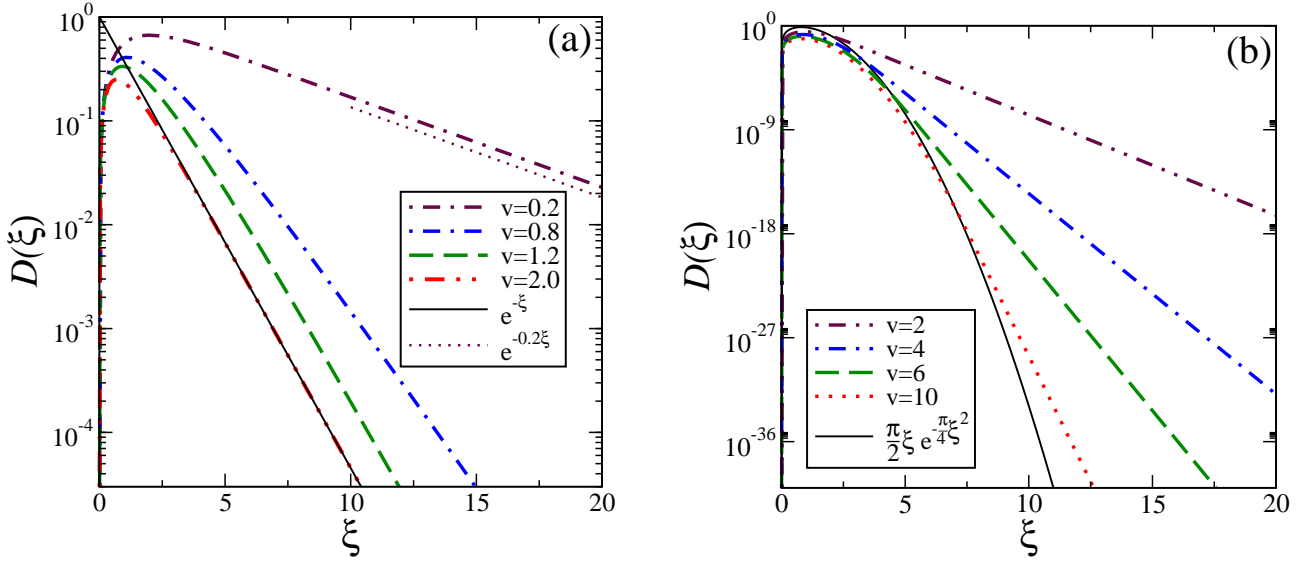

\centerline{\psfig{figure=durang4_ipdf_exp,width=3.2in,clip=} ~~~~
\psfig{figure=durang4_ipdf_erfc,width=3.2in,clip=}}
\caption[fig.ipdf.sans.input]{Stationary {\sc ipdf} $D(\xi,v)$
of the coagulation-diffusion process, as a function of the scaled interval
size $\xi$ and several values of the scaling variable $v$. Left panel (a): reset distribution $F(x)= \exp(-cx)$;
the {\sc ipdf} eq.~(\ref{4.4}) is shown for $v=[0.2,0.8,1.2,2.0]$ from top to bottom.
The distributions $e^{-\xi}$ and $e^{-0.2\xi}$
are also indicated.
Right panel (b): reset distribution $F(x) = \erfc(\demi\sqrt{\pi\,}\,cx)$; the {\sc ipdf} eq.~(\ref{4.5})
is shown with $v=[2,4,6,10]$ from top to bottom.
The distribution $\partial^2_{\xi}\; \erfc\left(\frac{\sqrt{\pi}}{2}\xi\right)= \frac{\pi}{2}\xi\e^{-\frac{\pi}{4}\xi^2}$ is also shown.
\label{fig.ipdf.sans.input}
}
\end{figure}

These functions are displayed in figure~\ref{fig.ipdf.sans.input}. First, in the left panel, the scaling function
$D_{(a)}$ is shown. For the simple coagulation-diffusion process under study here,
one would expect, consulting table~\ref{tab2}(b), a gaussian shape of the {\sc ipdf}. Clearly, this is no longer the
case in the presence of a reset. Rather, one sees that although $D(\xi)\stackrel{\xi\to 0}{\rar} 0$ as it should be
for a stationary {\sc ipdf} \cite{benA00},
for larger intervals one has an exponential distribution, typical of a system of {\em uncorrelated} particles.\footnote{This
is analogous to the finding of {\sc em} that the probability distribution of a random walk with reset is no longer gaussian.}
Furthermore, one observes that the effective density of particles in the large-$\xi$ regime (which can be read off
from the slope of $\ln D(\xi)$) depends in a non-trivial way on the scaling parameter $v=\alpha/c$. Namely, if
$v>1$, then the effective particle-density is simply unity, whereas if $v<1$,
that effective particle-density is equal to $v$
(it remains to be seen whether this kind of dynamical transition also occurs in different models).
In conclusion, the behaviour of the {\sc ipdf} at small scales is
quite distinct from the one seen at large scales.
This is a consequence of the fact that the stationary state in
the presence of a reset can no longer be described an equilibrium state.

A further aspect of this become apparent if a different reset configuration is analysed, see the right panel of
figure~\ref{fig.ipdf.sans.input} with the scaling function $D_{(b)}$.
Here, the reset is done to a configuration of particles as obtained from an usual
coagulation-diffusion process, with the natural correlations corresponding to a given density $c$.
At first sight, one might
expect that the reset to such a correlated configuration should lead to these correlations being maintained for all
interval sizes $\xi$. However, it can be seen from figure~\ref{fig.ipdf.sans.input} that this is not so. Rather,
in the stationary state the correlated particle configurations only describe the actual {\em stationary} {\sc ipdf}
only at small interval sizes $\xi$. At larger sizes, one observes again an effective distribution corresponding to
uncorrelated particles.

Intuitively, the observation from figure~\ref{fig.ipdf.sans.input}b may be understood as follows: through the reset rate $r$,
a further time scale $\tau_r\sim \alpha^{-2}$
is introduced which in turn creates a new length scale $\xi_{r}\sim \alpha$.
Between two reset events, the system
has on average enough time to reconstitute its natural
correlations up to scales $\xi\lesssim \xi_{r}$ but since
the resets are uncorrelated, beyond that scale its particles have become uncorrelated.
The non-equilibrium nature of
the stationary state manifests itself in strong short-distance correlations,
as prescribed by the original dynamics, and
an uncorrelated behaviour at large distances.

\subsection{{\sc ipdf} with input}

{}From the previous equation (\ref{eq.f}), the stationary ({\sc ipdf}) is cast in the scaling form
\bb
{\cal D}(x) := \frac{1}{2\rho_{stat}} \frac{\partial^2 f(x)}{\partial x^2}
= \frac{\beta^2}{\rho_{stat}} D(\beta x, c/\beta, \beta \mu)
\ee
with the three-variable scaling function
\bb \nn
D(\xi,u,y) &=& \frac{(\xi+y)\Ai(\xi+y)}{\Ai(y)}
- \pi y (\xi+y) \Ai(\xi+y) \frac{\Bi(y)}{\Ai(y)} \int_0^\infty \D Y \; F(uY/c)\Ai(Y+y)
\\
& & + \pi y (\xi+y) \Bi(\xi+y) \int_\xi^\infty \D Y \; F(uY/c)\Ai(Y+y)
\\ \nn
& & + \pi y (\xi+y) \Ai(\xi+y) \int_0^\xi \D Y \; F(uY/c)\Bi(Y+y)
\\ \nn
& & + \pi y F(u\xi/c)\left[\Bi(\xi+y)\Ai'(\xi+y) - \Bi'(\xi+y)\Ai(\xi+y)\right]
\ee
and the scaling variables $\xi:=\beta x$, $u :=c/\beta$ and $y:=\beta\mu$.

\begin{figure}[tb]
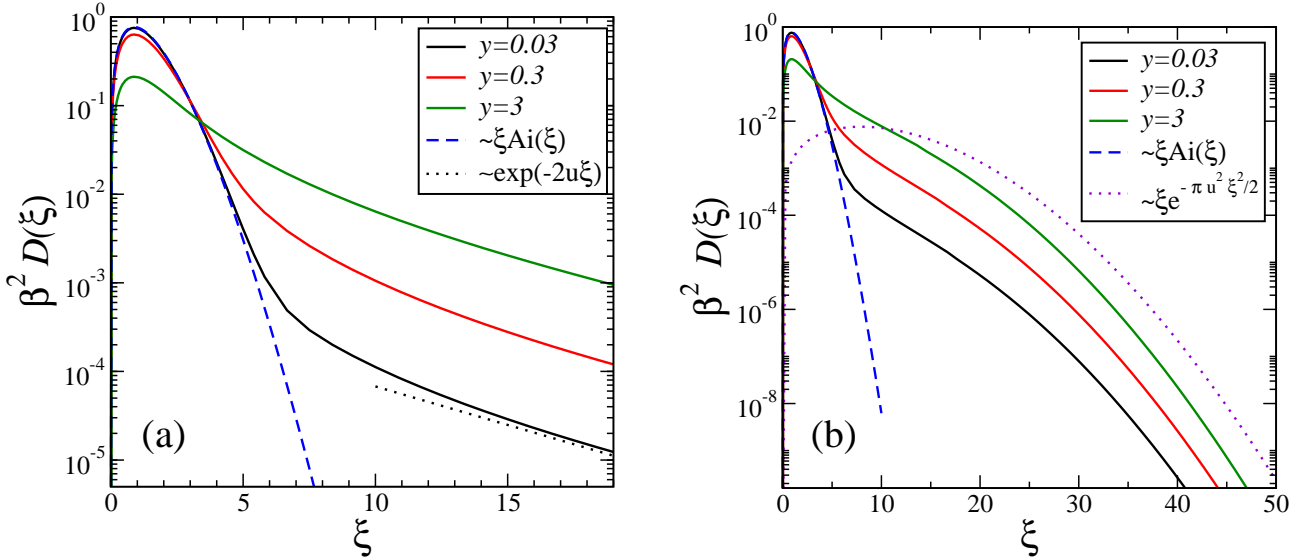

\centerline{\psfig{figure=durang4_ipdfinputexp,width=3.2in,clip=} ~~~~
\psfig{figure=durang4_ipdfinputerfc,width=3.2in,clip=}}
\caption[fig3]{Stationary {\sc ipdf} of the {\sc cdpr} with input, for the case $u=0.1$ and several values of $y$.
Left panel (a): reset distribution $F(x)= \exp(-cx)$. The {\sc ipdf}s for diffusion-coagulation with input, both for
input without a reset and for independent particles, are also shown.
Right panel (b): reset distribution $F(x) = \erfc(\demi\sqrt{\pi\,}cx)$.
The {\sc ipdf}s for diffusion-coagulation with input and for diffusion-coagulation are also shown.
\label{fig3}
}
\end{figure}

In figure~\ref{fig3}, the consequences of the reset are illustrated. First, for a reset to uncorrelated particles,
the behaviour seen in the left panel (figure~\ref{fig3}a)
is qualitatively the same as seen above in the case without input.
At short interval sizes, the system has enough
time between two resets to build up its natural correlations, so that the
shape of the {\sc ipdf} is essentially given by the
Airy function (see table~\ref{tab2}) and its stretched-exponential form.
For larger sizes, the particles become uncorrelated and the {\sc ipdf} goes over to a simple exponential.

A similar pattern is seen when resetting to configurations of correlated particles. In the right panel (figure~\ref{fig3}b),
the {\sc ipdf} for a reset to a configuration of simple diffusion-coagulation with mean density $c$ is shown.
With respect to the previous situation, the {\sc ipdf} of the reset distribution $F(\xi)$ falls off more rapidly for $\xi\gg 1$ than
the `natural' one of the underlying process. Yet, we see that the reset rate $r$ again sets a time scale such that for
sufficiently small interval sizes, the distribution of the empty intervals is the natural one of diffusion-coagulation with input and
goes over to the one put in by the reset for larger intervals.\footnote{We did not detect any evidence that for extremely large
values of $\xi$, the {\sc ipdf} would cross over to a simple exponential form.}

In any case, these examples illustrate the subtle nature of the stationary states in
simple particle-reaction models with a stochastic reset. The main new feature is a new scale set by the reset rate $r>0$, such that
at sufficiently small length scales, the `natural' correlations of the dynamics dominate whereas at larger length scales, those
of the reset configurations become dominant.

\section{Conclusions}

Analysis of the effects of a stochastic reset provides an alternative route to better appreciate the consequences
of the breaking of detailed balance. This breaking is required to obtain non-equilibrium stationary states. In the present work,
we have studied how the properties of a simple reaction-diffusion model are modified through the introduction of
a stochastic reset. This was achieved by identifying a new member in the class of models which may be solved exactly
through the `empty-interval method'.

A particular bonus of this technique is that it provides a very direct access to the distribution of the distances
between particles. In this way, we have seen that (i) the model's behaviour is not much affected by the reset at
short length scales but (ii) profoundly altered at larger scales. The coexistence of at least two kinds of correlations
at different length scales should be identified as the main mechanism which drives the system to a new non-equilibrium stationary
state.

Comparison of our analytical results with Monte Carlo simulations permit to identify how to set up analogous studies
in different many-body problems and/or networks, where exact analytical results may not be so readily available.

\appsektion{Remark on the Monte Carlo simulations}
In order to further illustrate the proper choice of a microscopic model of the {\sc cdpr},
which for continuous time would be described by
(\ref{1.2}), we compare two different choices for the transition rates in a Monte Carlo simulation:
\begin{figure}[tb]
\centering
\includegraphics[scale=0.4,angle=0]{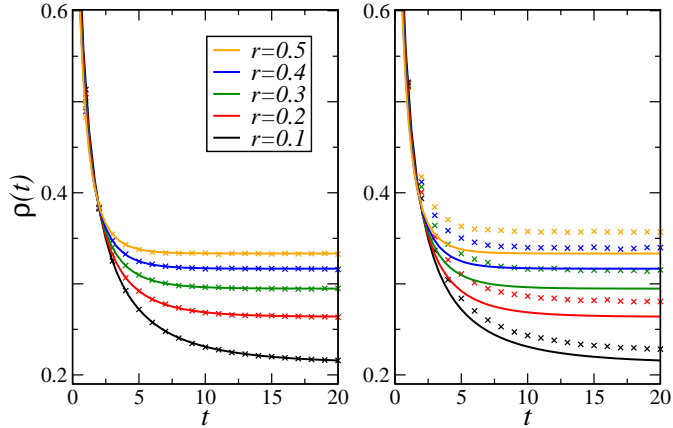}
\caption{Time-dependent particle-density $\rho(t)$
in two distinct Monte Carlo simulations on a periodic chain with ${\cal N}=512$ sites.
Left panel: result of method 1, right panel: results of method 2. The parameters used are $D=1/2$, $p=0.5$ and
$r=[0.1,0.2,0.3,0.4,0.5]$ from bottom to top. Initially the system was entirely filled.
The full lines give the exact analytical result, see section~2.
\label{abbA}}
\end{figure}
\begin{itemize}
\item \underline{\it Method 1 :} as already defined in section~1 in the main text.
We insist that the probability ${\cal P}_r$ chosen for the reset guarantees a reset with probability $r$ per sweep.
\item \underline{\it Method 2 :} for each sweep of $\cal N$ individual Monte Carlo steps,
the move to be carried out is globally chosen for all particles:
either ${\cal N}$ usual coagulation-diffusion steps are selected with probability $2D/(2D+r)$,
or else a global reset is chosen, with probability $r/(2D+r)$.
\end{itemize}
In figure~\ref{abbA}, the results for the time-dependent density $\rho(t)$ of these choices of the dynamics are shown,
for a periodic chain with ${\cal N}=512$ sites, $D=\demi$,
a reset to uncorrelated particles with mean density $p=0.5$ and several values of $r$.
Comparison with the exact result, derived in section~2,
shows that while the data obtained from method~1 perfectly agree, there is
no agreement with those from method~2.

\noindent
{\bf Acknowledgements:} This work was started during the 5$^{\rm th}$ KIAS conference on Statistical Physics.
We thank S.N. Majumdar and J.D. Noh
for useful discussions. {\sc mh} is grateful to KIAS Seoul for warm hospitality. This work
was partly supported by the Coll\`ege Doctoral franco-allemand Nancy-Leipzig-Coventry
(Syst\`emes complexes \`a l'\'equilibre et hors \'equilibre) of UFA-DFH
and also by the Basic Science Research Program through
the NRF Grant No.~2013R1A1A2A10009722.


\end{document}